\documentclass[preprint,aps,showpacs,nofootinbib,preprintnumbers,amsmath,amssymb]{revtex4-1}
\usepackage{epsfig}
\usepackage{bm}
\usepackage[usenames ,dvipsnames]{xcolor}
\usepackage{slashed}
\usepackage{caption}
\usepackage{subcaption}
\usepackage{graphicx,color}

\usepackage{multirow}
\begin{document}
\title{Searching for possible evidences of  new physics in $B\to V_1 V_2$ }

\author{C.Q. Geng$^{1,2,3}$ and  Chia-Wei Liu$^{1,2}$}
\affiliation{
	$^{1}$School of Fundamental Physics and Mathematical Sciences, Hangzhou Institute for Advanced Study, UCAS, Hangzhou 310024, China \\
	$^{2}$International Centre for Theoretical Physics Asia-Pacific, Beijing/Hangzhou, China \\
	$^{3}$Chongqing University of Posts \& Telecommunications, Chongqing 400065, China
}\date{\today}

\begin{abstract}
We explore  the decays of $B\to V_1V_2$ ($V_{1,2}= (\rho, \omega,K^*, \phi)$ and $B= (B^0, B^+,B_s)$) with transverse polarizations. 
We explicitly evaluate  the eigenstates of T-odd scalar operators involving spins for the first time, which offer physical insight among the T violating observables. 
Based on the helicity suppression of tree operators for transverse polarizations in the standard model (SM), 
we deduce that $\Delta \phi _p= \Delta \phi_\parallel - \Delta  \phi_\perp=0$ with $\Delta\phi_{\perp,\parallel}$  the weak phases of the transverse amplitudes.
In contrast, the experiments show that  $\Delta \phi _p (B^0 \to K^{*0} \omega)= -0.84\pm 0.54$, which could be  a signal of new physics.
There is also a discrepancy between our result in the SM and the experimental data for the transverse polarized branching ratio
 in $B^0 \to K^{*0} \omega$.
In addition, by counting the helicity flips,
 we argue that $\sin(\phi_\parallel - \phi_\perp ) \approx 0$,
 which meets well with most of the experiments except for the ones with $B^0\to K^{*0}\omega$, where $\phi_{\parallel,\perp}$ are the strong phases in the transverse amplitudes.
\end{abstract}
\maketitle

\section{Introduction}

It is known that the final state mesons in the $B$ decays of $B\to VT$
can be decomposed into three orthogonal states, where  $B= (B^0, B^+,B_s)\,,V= (\rho, \omega,K^*(892), \phi)$, and $T$  represent  spin-$n$\,($n\ge 1$) mesons. Although these decays suffer large theoretical uncertainties, they can be used to explore the helicity structures of the hadrons in the standard model~(SM). In particular, we are allowed to observe T violating observables originated from the triple momentum correlations~\cite{Valencia:1988it,Chiang:1999qn,Datta:2003mj,London:2004ws,LondonBtoVT,Gronau:2011cf,Kramer:1993yu}.

Note that among over a hundred measured parameters in the experiments for the $B$-meson decays~\cite{pdg},  
nearly forty  of them with more in the future are related to the T violating observables. Recently, $B^0\to K^{*0} \rho^0/\omega$ 
have been fully investigated for the first time by the LHCb Collaboration~\cite{Krho}.
 Particularly, the precision measurement in $B^0\to K^{*0} \rho^0$  is extraordinary, where the uncertainties of the polarized fractions are less than $4\%$. For instance, the data of LHCb has shown that~\cite{pdg,Krho} 
\begin{equation}\label{clue1}
f_L(B^0 \to K^{*0} \rho^0 ) = 0.173 \pm 0.026\,,
\end{equation}
where $f_L$ is the longitudinal polarized fraction.
In addition, the LHCb Collaboration has also updated the experimental data for $f_L(B_s \to \phi \phi)$ with the uncertainty is lower than
$2\%$~\cite{Bsphiphi}. Clearly, these polarization parameters grant us  new methods to test the SM and search for new physics~(NP)~\cite{Kagan:2004uw,Kagan:2004ia}. 

On the theoretical side, considerable progresses have been achieved~\cite{Suzuki:2001za,Bartsch:2008ps,Cheng:2008gxa,Cheng:2009xz,Cheng:2009cn,Zou:2015iwa,FQCDBu,FQCDBs,SU3,Yan:2018fif,FSI,Cheng:2001aa,Cheng:2009mu,Lu:2000hj,Lu:2005be,Li:2004mp,Li:2005hg,Beneke:2005we,Beneke:2006hg,Huang:2005if,Zhu:2005rt,Zhu:2005rx,Ali:2007ff,Bauer:1986bm,Ali:1997nh,Ali:1998gb,Ali:2007ff,Li:2004mp,Colangelo:2004rd,Cheng:2009ms,Raslan:2020dzh,Raslan:2020qch,Rui:2021kbn,Liu:2017cwl,Bell:2020qus,Lu:2006nza,Lu:2002ny,Cheng:2004ru,Ball:2006eu, Wang:2017rmh,Li:2021qiw,Jia:2021avh}. 
In the naive factorization, the amplitudes always contain $\langle V | (\overline{q}q')_{V\pm A}|0\rangle$, implying that 
$q$ and $\overline{q}'$ have opposite helicities
in $V$, and leading to $f_L\approx1$~\cite{Cheng:2001aa,Bauer:1986bm,Korner:1979ci}. 
It agrees well with the data in the decays dominated by tree diagrams. For instance, it is reported that 
$f_L(B^0\to \rho^+\rho^-)=0.990^{+0.021}_{-0.019}$~\cite{pdg}.
However, the argument is obviously invalid in non-factorizable processes. 
In 2003,
$f_L(B^0 \to \phi K^{*0})$ was measured around $0.5$ by BarBar~\cite{BaBar:2003spf} and Belle~\cite{Chen:2003jfa},
which made theorists confront the penguin annihilation type of diagrams~\cite{Li:2004mp,Colangelo:2004rd}, 
or solve the puzzle with  the charm-anticharm rescattering,
despite a large theoretical uncertainty~\cite{FSI,Cheng:2004ru}.

In  the perturbative QCD~(PQCD), the decays are calculated through six quark effective operators~\cite{Zou:2015iwa,Li:2021qiw}. 
Although the calculation 
can be done systematically with the heavy quark expansion, it suffers from the uncertainties in the hadron structure and QCD energy scale. On the other hand, in the improved QCD factrorization~(QCDF),  it is known that the effective Wilson coefficients depend on the helicity, which sheds light on the underlying complexity~\cite{Cheng:2009cn,Cheng:2008gxa}. Furthermore, in QCDF,  one is forced to add parameters by hand to overcome the endpoint divergence. Both frameworks have poor precisions in decay rates dominated by penguin operators, in which the uncertainties could be as large as $50\%$~\cite{Zou:2015iwa,Cheng:2009cn,Cheng:2009mu,FQCDBs,FQCDBu}. However, in PQCD, the uncertainties in $f_L$ are canceled due to the correlations. 
For instance, the latest study in PQCD gives that~\cite{Li:2021qiw}
\begin{equation}\label{clue2}
f_L(B^0\to K^{*0}\rho^0)  = 0.633^{+0.013}_{-0.012}\,^{+0.103}_{-0.099}\,^{+0.013}_{-0.015}\,,
\end{equation}
where the uncertainties are attributed to the shape parameters, Gegenbauer moments and the QCD energy scale, respectively.
The difference between Eqs.~\eqref{clue1} and \eqref{clue2} indicates that the underlying physics is not well understood yet.

On the kinematic aspect, as the vector mesons of $V_{1,2}$ in $B\to V_1V_2$ are polarized,
the decays 
 provide the possibility for observing the triple vector product (TVP) asymmetries.
In the literature, the decay amplitudes of  $B\to V_1V_2$ are expanded by the momenta and the spins. Explicitly,
 one has that~\cite{Valencia:1988it,Datta:2003mj}
\begin{equation}\label{newadded}
A\left( {B(p) \to V_1(p_1, \varepsilon_1) V_2(p_2, \varepsilon_2 )}\right) = a \varepsilon_1 \cdot \varepsilon_2 + b (p \cdot \varepsilon_1 ) ( p \cdot \varepsilon_2) + i \frac{c}{m_B^2 }\epsilon_{\mu \nu \rho \sigma} p^{\mu} (p_1^\nu - p_2 ^\nu ) \varepsilon_1^{*\rho}\varepsilon_2^{*\sigma}\,,
\end{equation}
where $A$ stands for the amplitude, $a$, $b$ and $c$ correspond to the $s$, $d$ and $p$ wave amplitudes, 
 $p_{i}$ and  $\varepsilon_{i}$ ($i=1,2$)  are the 4-momenta and   polarized vectors for the vector mesons of $V_i$, respectively, and  $\epsilon_{\mu\nu\rho\sigma}$ is the total antisymmetric tensor.
Without the final state interactions, the T symmetry would demand $a$, $b$ and $c$ to be real. 
As the term associated with  $c$ in Eq.~(\ref{newadded})
takes the TVP form in the rest frame of the $B$ meson,
the results based on $\Im(ac^*)$ and $\Im(bc^*)$ are referred to as the ``triple product correlations'' in the literature.
However, the spins~$(\varepsilon_{1,2})$ do not commute with the TVP. To be explicit, we have~\cite{spinoperators}
\begin{equation}
[(\vec{s}_1 \times \vec{s}_2 ) \cdot \hat{p}_1, \vec{s}_1] = i (\vec{s}_1 \cdot \hat{p}_1) \vec{s}_2 - i(\vec{s}_1 \cdot\vec{s}_2 )\hat{p} _1 \,,
\end{equation} 
 where the subscripts label the particles, while $\vec{s}$ and $\hat{p}$ are the spin and the norm of the three-momentum operators, respectively. 
Although $\Im(ac^*)$ and $\Im(bc^*)$ are related to the T violating effects, it is still questionable to discuss the TVP with Eq.~\eqref{newadded}, and   a reanalysis on 
the TVP asymmetry is therefore required. To this end, we expand the states with T-odd scalar operators ($\hat{T}$), to naturally demonstrate the T violating observables.

 This work is organized as follows. In Sec.~\MakeUppercase{\romannumeral 2}, we  construct  T-odd observables for $B\to VT$. In Sec.~\MakeUppercase{\romannumeral 3}, we discuss the numerical results of $B\to V_1V_2$.
 We present the summary in Sec.~\MakeUppercase{\romannumeral 4}.

\section{T-odd observables}
 In this section, we study the possible T-odd observables in $B\to VT$.
For a given  T-odd operator, $\hat{T}_i$, one can define an asymmetry parameter as 
\begin{eqnarray}\label{TNaive}
\Delta _{t_i}\equiv \frac{\Gamma(\lambda_{t_i}>0)  -\Gamma(-\lambda_{t_i}<0)   }{\Gamma(\lambda_{t_i}>0) +\Gamma(-\lambda_{t_i}<0) } = 
\frac{|A(\lambda_{t_i}>0)|^2  -|A(-\lambda_{t_i}<0)|^2   }{|A(\lambda_{t_i}>0)|^2 +|A(-\lambda_{t_i}<0)|^2} 
\end{eqnarray}
where $\lambda_{t_i}$ is the eigenvalue of $\hat{T}_i$, and 
 $\Gamma(\lambda_{t_i})$ and $A(\lambda_{t_i})$ represent the partial decay rate and amplitude for $B\to \lambda_{t_i}$, respectively.
Here, we have chosen 
$\lambda_{t_i}>0$ to avoid the ambiguity.
If the relevant effective Hamiltonian for the decays
is invariant under the time reversal transformation, we have~\cite{GroupTheory}
\begin{equation}\label{newadd}
|\langle \lambda_i; {\text{``out''}} | {\cal H}_{eff} | B\rangle |^2 - |\langle -\lambda_i; {\text{``in''}} | {\cal H}_{eff}  | B\rangle |^2 = 0 \,,
\end{equation}
where ``out''~(``in'') denotes $t \to \infty (-\infty)$. 
Note that if the final state interactions~(FSIs) are absent, ``out'' and ``in''
can be interchanged freely, and the left side of Eq.~\eqref{newadd} is proportional to the numerator in the right side of  
Eq.~\eqref{TNaive}, resulting in that $\Delta_{t_i}$ is a T-odd observable.

Clearly, to calculate $\Delta_{t_i}$, one has to explicitly obtain the eigenstates of $\hat{T}_i$. To do so, it is important to choose a commuting set of operators to describe the states as emphasized in the previous section. In the final states, the angular momenta are always constrained by the parent particles, so it is preferable to describe the final states with $J^2$ and $J_z$.

To be compatible with $J^2$ and $J_z$, we choose $\hat{T}_i$ as an $SO(3)$ rotational scalar, which automatically commutes with the angular momenta. In this study, we consider three of the most simple $\hat{T}_i$, given as 
 \begin{eqnarray}\label{definitionforTi}
 \hat{T_2}&=&(\vec{s}_1\times \vec{s}_2)\cdot \hat{p}_1\,,~~~~~~~
\hat{T}_3=(\vec{s}_1\cdot\hat{p}_1)\hat{T}_2
+\hat{T}_2(\vec{s}_1\cdot\hat{p}_1)\,,\nonumber\\
\hat{T}_4 &=& (\vec{s}_1\cdot\vec{s}_2)\hat{T}_2
+\hat{T}_2(\vec{s}_1\cdot\vec{s}_2)-
(\hat{p}_1\cdot\vec{s}_1)(\hat{p}_1\cdot\vec{s}_2)\hat{T}_2
-\hat{T}_2(\hat{p}_1\cdot\vec{s}_1)(\hat{p}_1\cdot\vec{s}_2)\,,
\end{eqnarray}
where the subscripts in $\hat{T}_i$ denote the numbers of spin operators. The eigenstates can be directly computed and expressed with the helicity states as demonstrated in Appendix A. Notably, we find that $\lambda_{t_i}$ are discrete, in contrast to Ref.~\cite{Valencia:1988it}, which treats $\lambda_{t_i}$ as continuous variables.

Plugging Eqs.~(\ref{lambdat2}),  (\ref{lambat3})  and (\ref{lambdat4}) in Eq.~\eqref{TNaive},  we obtain
\begin{eqnarray}\label{consistent}
\Delta_{t_2} &=& \frac{2\sqrt{f_Lf_\perp} \sin (\phi^B_\perp)}{f_L +f_\perp}\,,\nonumber\\
\Delta_{t_3} &=& \frac{2\sqrt{f_L f_0 } \sin (\phi^B_\parallel)}{f_L + f_0}\,,\nonumber\\
\Delta_{t_4} &=& \frac{-2\sqrt{f_\perp f_\parallel } \sin (\phi_\parallel^B-\phi_\perp^B)}{f_\perp + f_\parallel}\,,
\end{eqnarray}
where the superscript of ``$B$'' denotes the parent $B$-meson, and
\begin{eqnarray}
f_i = \frac{| A (H_i)|^2}{\sum_{j=0,\pm} |A(H_j)|^2}\,,
\end{eqnarray}
where the definitions of $H_i$ are given in Eq.~\eqref{Hpm}, and  $\phi_{\perp,\parallel}$ correspond to the  phases between $A(H_{\perp, \parallel} )$ and $A(H_0)$.

Since the oscillation between $|\pm \lambda_i\rangle$ is possible via the final state interactions, to obtain a true T violating observable, we have to subtract the effect of the final state interactions from the charge conjugated 
part.
Under the parity,  $\hat{T_i}$ transform according to the number of spin operators, which are given as
\begin{equation}
I_s^\dagger \hat{T}_iI_s = (-1)^{i-1} \hat{T}_i\,,
\end{equation}
resulting in
\begin{equation}
 \hat{T}_i I_s |\lambda_{t_i}\rangle =(-1)^{i-1}\lambda_{t_i} I_s |\lambda_{t_i}\rangle\,,
\end{equation}
where $I_s$ is the parity operator. Accordingly, if CP is conserved, the decay widths are related to the charge conjugated ones as 
\begin{equation}
\Gamma (\lambda_{t_i}) =\overline{\Gamma}\left((-1)^{i-1}\lambda_{t_i}\right)\,.
\end{equation}
Consequently, the T or CP asymmetries are given by 
\begin{eqnarray}
\label{CPA}
\Delta^{CP}_{t_i}=
\Delta_{t_i} +(-1)^{i} \,\overline{\Delta}_{t_i} \,,
\end{eqnarray}
where the overline denotes the charge conjugation.
	Note that strong phases are not needed to have a non-zero value of the asymmetry in Eq.~(\ref{CPA}). In terms of the complex phases, the CP-odd parameters are given as
\begin{eqnarray}
\Delta \phi_\perp = (\phi_\perp^{\overline{B}} -\phi_\perp^{B} \pm \pi)/2 \,,\,\,\,\,\,\,\Delta\phi_\parallel =(\phi_\parallel^{\overline{B}}-\phi_\parallel^{B})/2\,,\,\,\,\,\,\,\Delta\phi_p = \Delta\phi_\parallel - \Delta\phi_\perp\,,
\end{eqnarray}
which are related to $\Delta_{t_2}^{CP}$, $\Delta_{t_3}^{CP}$  and $\Delta_{t_4}^{CP}$, respectively. Here, $\pm\pi$ in $\Delta \phi_\perp$ take account the fact that $\Delta_{t_2}+\overline{\Delta}_{t_2}=0$ if CP is conserved. The signs are chosen such that $\pi/2 > \Delta \phi_\perp > -\pi/2$.

 In the literature, $\Delta \phi_\parallel$ and $\Delta \phi_\perp$ are the so-called “true” CP violating phases, whereas the “fake” ones are given by
\begin{equation}
\phi_\perp =\phi_\perp^B + \Delta \phi_\perp\,,\,\,\,\,\,\,\phi_\parallel =(\phi_\parallel^{\overline{B}}+\phi_\parallel^{B})/2\,,\,\,\,\,\,\,\phi_p = \phi_\parallel - \phi_\perp\,,
\end{equation}
which are originated by the strong interaction. Likewise, we define that
\begin{equation}
\Delta^S_{t_i} = \Delta_{t_i} - (-1)^i \overline{\Delta}_{t_i}\,,
\end{equation}
which vanish if the final state interaction and CP violation are both absent.
Clearly, non-zero values of $\Delta^S_{t_i}$ do not necessarily  imply CP or T violation as they can purely arise from the final state interactions.
 
In the experiments, the information of T-odd observables is often extracted via the cascade decays with four final state particles, namely $B \to V_1(\to 0^- 0^-)T_2(\to 0^-0^-)$, represented by the triple momentum correlation of $(\vec{p}_1 \times \vec{p}_2)\cdot \vec{p}_3$, which is a P-odd operator, 
and therefore  $\Delta_{t_3}$, 
which is P-even, is not expected to be observed. On the other hand, if the final states of the cascade decays involve more than four particles, it is possible to construct P-even quantities with the form $(\vec{L}_{1,2}\times \vec{p}_3)\cdot \vec{p}_4$ related to $\Delta_{t_3}$, where $\vec{L}_{1,2}$  are the orbital angular momentum operators formed by two particles. The argument is affirmed by the explicit forms of the angular momentum distributions, where $\sin(\phi_\parallel)$ appears only in the final states with more than four particles~\cite{LondonBtoVT}.

Since $\hat{T}_4$ involves  interchanging the helicity twice (see Eq.~\eqref{C4}), one naturally expects that 
\begin{equation}\label{co}
\Delta^S_{t_4}\approx 0 ,
\end{equation}
which will be taken as an assumption and checked on its validity throughout this work.
In the decays with $f_\perp \approx f_\parallel$, 
such as those of  $B\to V_1 V_2$,
it means that
\begin{equation}\label{cornerstone}
\sin(\phi_p)\approx 0\,.
\end{equation}
The available  data for $B \to V_1T_2$ from the particle data group~(PDG)~\cite{pdg}  are collected in Table~\ref{data}.   The correlations between $\phi_\parallel$ and $\phi_\perp$ in the experiments are assumed to be zero, unless explicitly given by the data.

It is known that  the decay parameters, obtained from the angular distributions, exist two-fold solutions, which are related by~\cite{Chiang:1999qn}
\begin{eqnarray}
\phi_\perp &\rightarrow&  \pi -\phi_\perp\,,\nonumber\\
\phi_\parallel &\rightarrow& -\phi_\parallel\,.
\end{eqnarray}
These two solutions have very different meanings, since the values of $\phi_p = 0$ and $ \pi$ correspond to $H_- = 0$ and $H_+=0$, respectively.
In $B^0\to K^{*0}(892)\rho^0/\omega$, we have adopted a different set of solutions compared to the original one~\cite{Krho}
with the reason explained in the next section.
\begin{table}[t]
	\captionsetup{justification=raggedright,
		singlelinecheck=false
	}
	\caption{ Data from the PDG~\cite{pdg}, where we have adopted a different set of solution compared to the original one for the decays of $B^0\to K^{*0}(892)\rho^0/\omega$~\cite{Krho}, related as $\phi_\perp\to \pi -\phi_\perp$ and $\phi_\parallel \to -\phi_\parallel $.}
	\begin{tabular}{lcccc}
		\hline
		channel&$\phi_\perp$& $\phi_p$&$\Delta \phi_\perp$&$\Delta\phi_p$\\ 
		\hline
		$B^+\to \phi K^{*}(892)^+$~\cite{Aubert:2007ac,Chen:2005zv}&$2.58\pm 0.17$&$-0.22\pm0.16$&$-0.19\pm0.21$&$0.12\pm0.19$\\
		$B^0\to  K^{*}(892)^0\rho^0 $~\cite{Krho} & $0.79\pm 0.11$&$-0.018\pm0.056$&$0.123\pm0.092$&$-0.014\pm0.043$\\
		$B^0\to  K^{*}(892)^0\omega $~\cite{Krho}&$0.7\pm0.9$&$ 0.3 \pm 0.9$&$0.28\pm 0.56$&$-0.84\pm 0.54$\\
		$B^0\to \phi K^{*}(892)^0$~\cite{pdg}&$2.53\pm0.09$&$-0.10\pm 0.14$&$0.08\pm 0.05$&$-0.03\pm0.07$\\
		$B^0 \to J/\psi K^{*}(892)^0$~\cite{pdg} & $2.96\pm 0.05$&$-5.88 \pm 0.06$&-&-\\
		$B^0\to\phi K^{*}_2(1430)^0$~\cite{pdg}&$4.5\pm 0.4$ &$-0.5\pm 0.6$&$-0.2\pm 0.4$&$-0.7\pm 0.6$\\
		$B^0 \to \psi (2s) K^{*0}(892)$~\cite{xc1} & $2.8\pm 0.3$ &$-5.6\pm 0.5$&-&-\\
		$B_s \to K^{*}(892)^0 \overline{K}^{*}(892)^0$~\cite{BsKK}&$2.62\pm 0.69$&$-0.22\pm 0.77$&-&-\\
		$B_s \to \phi \phi $~\cite{Bsphiphi}&$2.56\pm 0.06$&$-0.26 \pm 0.18$&$0.044\pm 0.062$&$-0.030\pm 0.084$\\
		$B_s\to J/\psi \phi$~\cite{pdg}& $2.9\pm0.4$&$0.2\pm 0.4$&-&-\\
		$B_s\to \psi(2s) \phi $~\cite{pdg}&$3.3\pm 0.4$&$0.4\pm0.4$&-&-\\
		$B_s \to J/\psi\phi$~\cite{Aaij:2021mus}&$2.41^{+0.44}_{-0.43}$&$0.70^{+0.43}_{-0.44}$&\\
		\hline
	\end{tabular}
\label{data}
\end{table}

While $\phi_\perp$ can be in general unbounded, $\phi_p$ in Eq.~\eqref{cornerstone} is consistent with the data except for the decay channels involving $c\overline{c}$. In these cases, the interacting periods for the final states are longer, due to the smaller velocity. 
 In the future, $\Delta_{t_4}^S\approx 0$ can be a useful benchmark to test the theory. In fact, our ansatz is well consistent with the theoretical approaches in the literature~\cite{Zou:2015iwa,FQCDBs,FQCDBu,SU3,Beneke:2006hg,Wang:2017rmh}. 

Likewise, for  the  $B$ decay with a pair of spin-2 mesons we find that 
\begin{eqnarray}
\sin(\phi_{\perp 1} - \phi_{\parallel 1}) &\approx &0\,, \nonumber\\
\sin(\phi_{\perp 2}-\phi_0)&\approx& 0\,.
\end{eqnarray} Here,
$\phi_\lambda$ are the strong complex phases of  $A(H_\lambda)$ and 
\begin{eqnarray}
A(H_{\parallel_i,\perp_i}) = \frac{1}{\sqrt{2}}\left[
A(H_{i}) \pm A(H_{-i})
\right]\,,
\end{eqnarray}
where $i=0,1,2$ denote the helicities.
 As there has been  only one experiment   with a large  uncertainty so far~\cite{BsKK}, it is hard to   draw any conclusion at this stage.

\section{$B\to V_1V_2$ decays}\label{section3}
In this section, we would like to explore the consequences of the ansatz in Eq.~\eqref{co} with the $SU(3)$ flavor~$(SU(3)_F)$ symmetry. To this end,
we concentrate on  the decays of $B\to V_1V_2$.
For  simplicity, we do not explicitly write down the meson masses.
We begin our study with  the constituent quark approach, in which  the spins of vector mesons are originated from the quark spins.
Accordingly, $H_+$ requires the quarks and antiquarks  to have  positive helicities, while $H_{-}$  negative ones.  
Since the decays of $\overline{b}$ release a great amount of energy, the chirality is related to the helicity.
Explicitly, it means that $H_\pm$ contain a pair of $\overline{q}_{L(R)}$ and $q_{R(L)}$.

The effective Hamiltonian for the charmless $b$ decays is given by~\cite{Buchalla:1995vs}
\begin{equation}
\mathcal{H}_{\mathrm{eff}}=\frac{G_{F}}{\sqrt{2}}\sum_{f=d,s}\left[V_{ub}^*V_{uf}(C_{1} O_{1}^{f}+C_{2} O_{2}^{f})
-V_{tb}^*V_{tf} \left(\sum_{i=3}^{10} C_{i} O_{i}^f + C_{7\gamma} O^f_{7\gamma} + C_{8G} O_{8G}^f
\right)
\right]\,,
\end{equation}
with
\begin{eqnarray}
&&O_1^f = (\overline{b}_\alpha u_\beta)_L (\overline{u}_\beta f_\alpha)_L\,,\,\,\,\,O_2^f = (\overline{b}_\alpha u_\alpha)_L (\overline{u}_\beta f_\beta)_L\,,\nonumber\\
&&O_3^f = (\overline{b}_\alpha f_\alpha)_L \sum_{q} (\overline{q}_\beta q_\beta)_L\,,\,\,\,\,O_4^f = (\overline{b}_\alpha f_\beta)_L \sum_{q}(\overline{q}_\beta q_\alpha)_L\,,\nonumber\\
&&O_5^f = (\overline{b}_\alpha f_\alpha)_L \sum_{q}(\overline{q}_\beta q_\beta)_R\,,\,\,\,\,O_6^f = (\overline{b}_\alpha f_\beta)_L \sum_{q}(\overline{q}_\beta q_\alpha)_R\,,\nonumber\\
&&O_7^f = \frac{3}{2}(\overline{b}_\alpha f_\alpha)_L \sum_{q}e_{q}(\overline{q}_\beta q_\beta)_R\,,\,\,\,\,O_8^f =\frac{3}{2} (\overline{b}_\alpha f_\beta)_L \sum_{q}e_{q}(\overline{q}_\beta q_\alpha)_R\,,\nonumber\\
&&O_9^f = \frac{3}{2}(\overline{b}_\alpha f_\alpha)_L \sum_{q}e_{q} (\overline{q}_\beta q_\beta)_L\,,\,\,\,\,O_{10}^f =\frac{3}{2} (\overline{b}_\alpha f_\beta)_L \sum_{q}e_{q}(\overline{q}_\beta q_\alpha)_L\,,\nonumber\\
&&O^f_{7\gamma} = \frac{e}{4 \pi^{2}} m_{\mathrm{b}} \bar{b}_{\alpha} \sigma^{\mu \nu}f_{\alpha L} F_{\mu \nu}\,,\,\,\,\,O_{8 G}=\frac{g}{4 \pi^{2}} m_{\mathrm{b}} \bar{b}_{\alpha} \sigma^{\mu \nu} T_{\alpha \beta}^{a} f_{\beta L} G_{\mu \nu}^{a}
\end{eqnarray}
where $q=(u\,, d \,, s\,,c)$, $G_F$ stands for the Fermi constant, $C_i$ represent the Wilson coefficients, $\alpha$ and $\beta$ correspond to the color indexes,
 $O_{i(j)}$ with $i$=3,4,5,6~($j$=7,8,9,10) are the so-called QCD penguin (electroweak pengiun) operators, 
the subscript of $L(R)$ implies the current structure of $V-A~(V+A)$, $O_{7\gamma}$ and $O_{8G}$ are named as magnetic penguin operators, $F_{\mu\nu}~(G_{\mu\nu}^a)$ is the photon~(gluon) field strength tensor, $f_{\alpha L}~(f_\alpha=d_\alpha$ or $ s_\alpha)$ is the left-handed quark,  $m_b$ is the $b$ quark mass, and $T^a$ are the Gell-Mann matrices.
Note that $\sum_{q}\overline{q}_\alpha q_\beta$ is  an $SU(3)_F$ flavor singlet, which gives the  simple flavor structure for the QCD penguin operators. 
The operators $O_{1,2,3,4,9,10}$ produce at least one $q_{\alpha L}$ and one $\overline{q}'_{\beta L}$ in final states, which are suppressed by the helicity flipping in $A(H_\pm)$. On the other hand, $O_{5,6,7,8}$ can contribute to either $A(H_0)$ or $A(H_+)$. Furthermore, due to the hierarchy of $C_{7,8}/C_{5,6}\approx \alpha_{em} $ with $\alpha_{em}$ the fine structure constant,  the electroweak operators  $O_{7,8}$  can also be safely neglected~\cite{Buchalla:1995vs}.

Since $f_{\alpha L}$ is left-handed and only the transverse photon operators are physical, $O_{7\gamma}$ only contributes to $A(H_+)$~\cite{Beneke:2005we}. The leading order amplitudes are factorizable, described by $B\to V_1$ and $\gamma \to V_2$,   given as~\cite{Beneke:2005we}
\begin{equation}\label{EDMAmp}
A(P_{7\gamma}) =-i e^{i\delta_{1,2}} \frac{\text{G}_F}{\sqrt{2}}V_{tb}^*V_{tf} m_b T_1^{V_1} f_{V_2} a'_{V_1} a_{V_2} {\cal C}'_{7\gamma}  \frac{4\alpha_{em}}{3\pi }\frac{m_B^2 }{m_{V_2}}\,,
\end{equation}
where $\delta_{1,2}$ are the strong phases,
 $T_1^{V_1}\approx 0.4$ is the tensor form factor~\cite{Lu:2002ny}, $f_{V_2} $ is the $V_2$-meson decay constant,
 ${\cal C}_{7\gamma}'$ represents the effective Wilson coefficient including  spectator-scattering effects~\cite{Beneke:2001at},  and
  $a_{V_1}'$ and $a_{V_2}$ correspond to the overlapping factors for $B \to V_1$ and $\gamma \to V_2$ 
with their explicit values, given by~\cite{Lu:2006nza}
\begin{eqnarray}
&&a_{\rho^+,K^{*+}, K^*0,\overline{K}^{*0}, \phi}'= 1\,,\,\,\,\,a_{\omega}'=-a_{\rho^0}'  = \frac{1}{\sqrt{2}}\,,\nonumber\\
&&a_{\rho^0} = \frac{3}{2\sqrt{2}}\,,\,\,\,\,a_{\omega} = \frac{1}{2\sqrt{2}}\,,\,\,\,\,a_\phi = - \frac{1}{2}\,,
\end{eqnarray}
respectively.

In short distance contributions, we conclude that $A(H_+)$ is dominated by $O_{5,6,7\gamma, 8G}$, whereas $A(H_{-})$ is highly suppressed, which agrees with the analyses of PQCD and QCDF~\cite{Zou:2015iwa,Cheng:2009cn,Beneke:2005we}.
It is important to emphasize  that the above argument is only valid for short distance physics. $A(H_+)$ may potentially oscillate to $A(H_{-})$, via the complex phases between $A(H_\perp)$ and $A(H_\parallel)$, induced by the final state interactions~\cite{Suzuki:2001za}. However, by demanding Eq.~\eqref{cornerstone}, we find that the oscillation is forbidden and $A(H_{-})$ remains suppressed.  This is the reason that we choose a different set of solutions for $B^0 \to K^{*0} \rho^0/\omega$ compared to that in  Ref.~\cite{Krho} (see Table~\ref{data}).  The direct result with $A(H_-)=0$ is given by
\begin{equation}\label{df}
\delta f \equiv f_\parallel - f_\perp = 0\,	.
\end{equation}
The available experimental data are shown in Table~\ref{fitting}.  We see that Eq.~\eqref{df} meets well with  the current measurements.
In the following, we use $A(H_-)=0$ for a quantitative analysis. 

In our numerical calculation,
we  focus on the branching ratios with  positive helicities
and take  that
$(f_\phi,f_\omega,f_\rho) = (215\pm 5,187\pm5,216\pm 3)$~MeV~\cite{Ball:2006eu,Ball:2004rg} and $|{\cal C}'_{7\gamma}|^2 = 1.65^{+0.018}_{-0.017}$~\cite{Beneke:2001at}. The amplitudes from $O_{5,6}$ are extracted from the experimental values with the $SU(3)_F$ symmetry as described by the following.
Due to the simple flavor structure in the penguin operators, there are only few undetermined parameters.
 The possible diagrams are illustrated in Fig.~1, where 
  $q$ represents the sum of $(u,d,s)$, and 
 $\otimes(\odot)$ stands for the coupling of the left (right)-handed current.
Figs.~1c and 1d involve a right-handed  quark and antiquark in the final states,
so they would contribute to $A(H_0)$ only.
We are left with Figs.~1a and 1b, representing the penguin annihilation~($P_A$)  and flavor singlet penguin annihilation~($P_S$), respectively. 
The  amplitudes of $A[P_A(\overline{b} \to \overline{d})]$  and $A[P_A(\overline{b} \to \overline{s})]$ 
associated with Fig.~1a
  are related by
\begin{equation}\label{SysH}
p_A \equiv A[P_A( \overline{b} \to \overline{d} )]/V_{td} =A[P_A( \overline{b} \to \overline{s} )]/V_{ts} \,,
\end{equation}
where we only consider those with positive helicity.
Similarly, we have
\begin{equation}\label{SysH2}
p_S \equiv A[P_S( \overline{b} \to \overline{d} )]/V_{td} =A[P_S( \overline{b} \to \overline{s} )]/V_{ts} \,.
\end{equation}
From  Eqs.~\eqref{SysH} and \eqref{SysH2}, it is straightforward to see that the  
amplitudes of $\overline{b} \to \overline{d}$ are  relatively suppressed by $|V_{td}/V_{ts}| = 0.210\pm 0.008$~\cite{pdg}. 
With the $SU(3)_F$ symmetry, 
the decay amplitudes  can be parametrized with $p_A$ and $p_S$, given in the second columns of  Tables~III and III, 
in which $V_{ts}$~$(V_{td})$ has not been shown explicitly for $\overline{b}\to \overline{s} $~$(\overline{b}\to \overline{d} )$.

Note that though the contributions from $O_{8_G}$ are not calculated as $O_{7_\gamma}$, still, we can parametrize it with the $SU(3)_F$ structure.
We note that
in contrast to $O_{7_\gamma}$, in which the photon operator would inevitably break the $SU(3)_F$ symmetry, the gluon magnetic operators in ${\cal H}_{eff}$ share the same $SU(3)_F$ structure with the QCD penguin ones, and therefore
 the contributions of $O_{8_G}$ can be absorbed into $A(P_A)$ and $A(P_S)$, as long as the $SU(3)_F$ symmetry is reliable.

\begin{figure}
		\captionsetup{justification=raggedright,
		singlelinecheck=false
	}
	\begin{subfigure}{.4\textwidth}
		\centering
		\includegraphics[width=.8\linewidth]{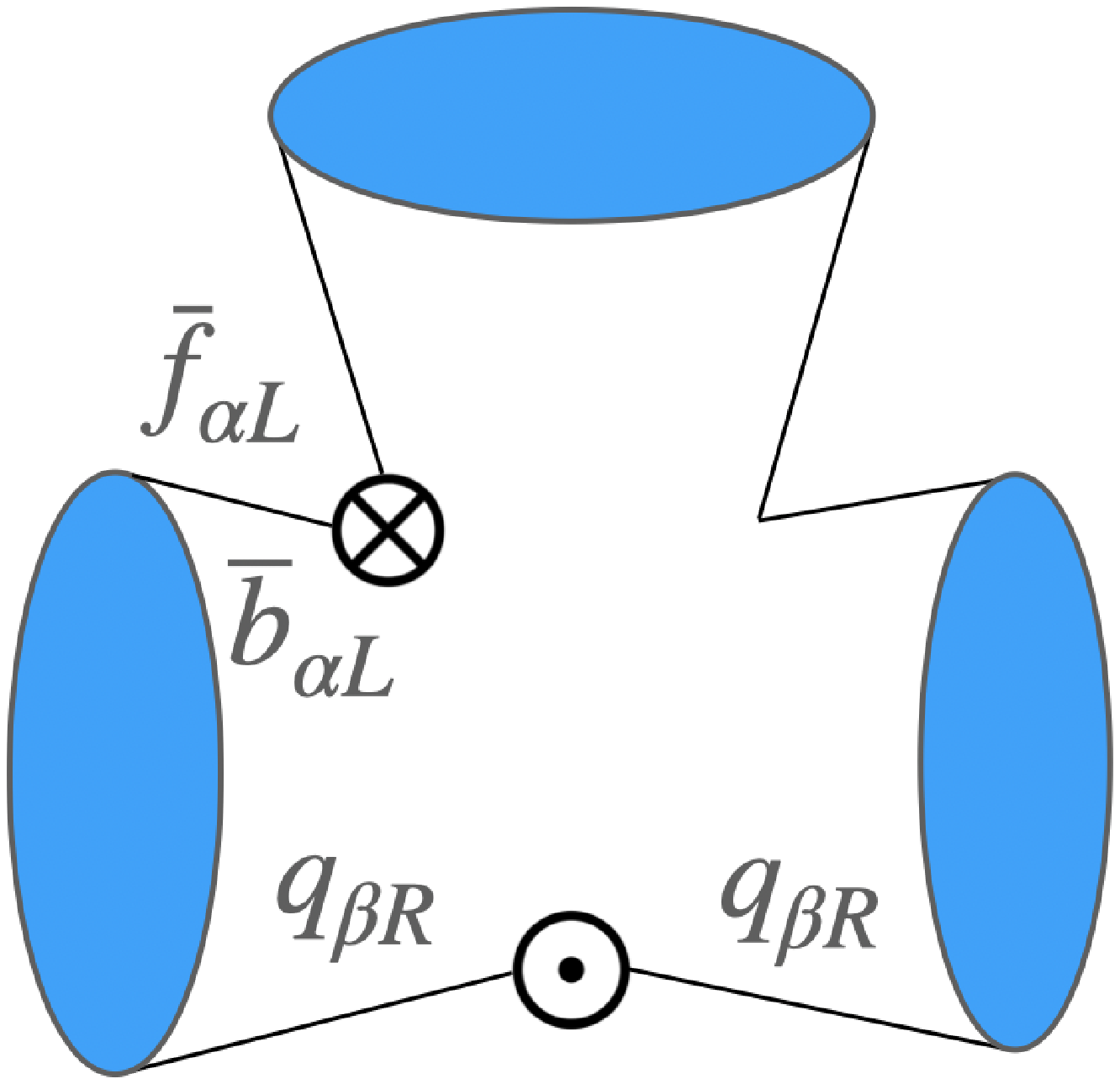}  
		\caption{}
	\end{subfigure}
	\begin{subfigure}{.4\textwidth}
		\centering
		\includegraphics[width=.8\linewidth]{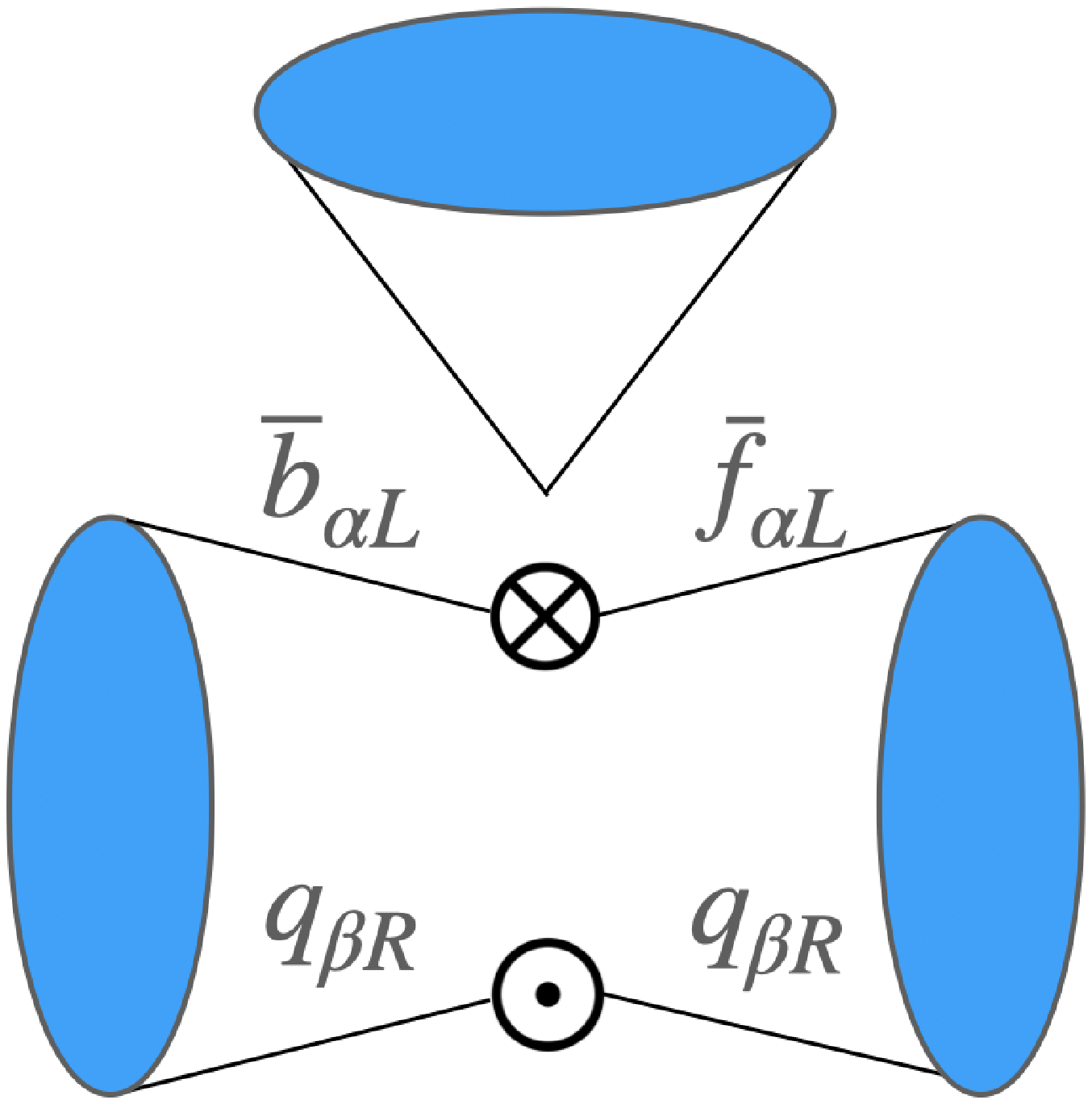}  
		\caption{}
	\end{subfigure}
	\newline
	
	\begin{subfigure}{.4\textwidth}
		\centering
		\includegraphics[width=.8\linewidth]{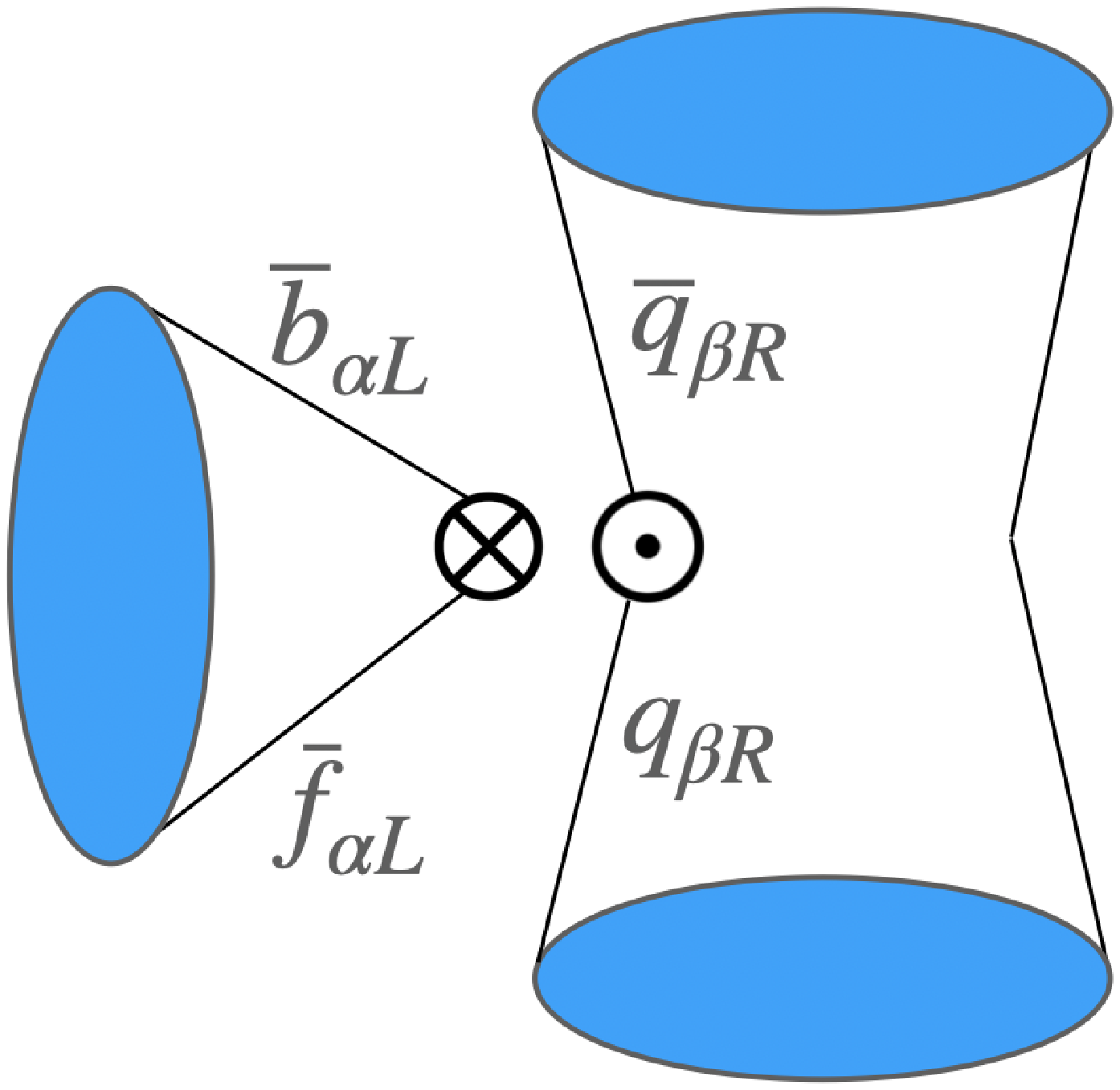}  
		\caption{}
	\end{subfigure}
	\begin{subfigure}{.4\textwidth}
		\centering
		\includegraphics[width=.8\linewidth]{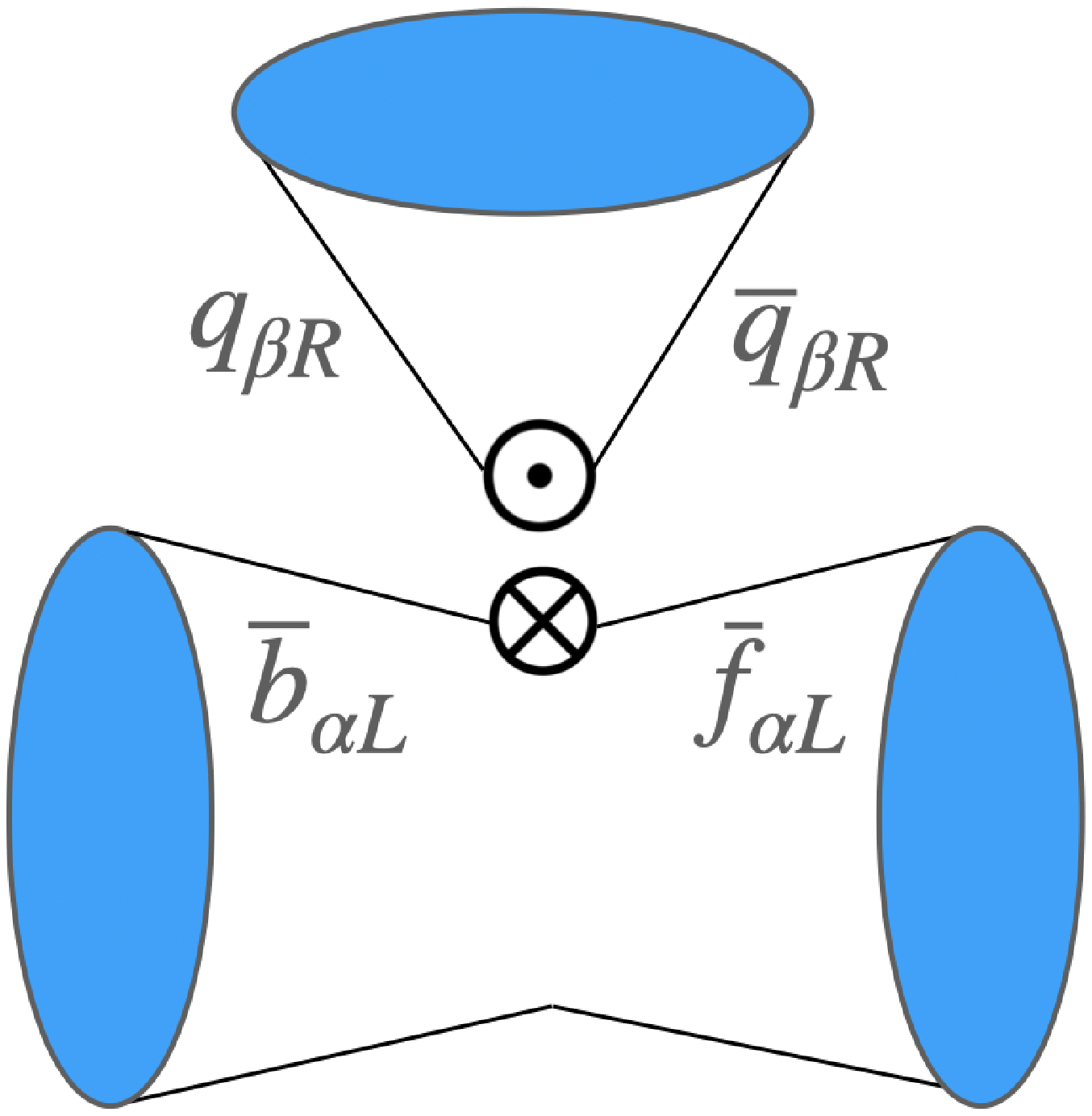}  
		\caption{}
	\end{subfigure}
	\caption{
		Topological diagrams for penguin operators $O_5$ and $O_6$, where (a), (b) and (c) correspond to the penguin annihilations, while (d)  the penguin induced tree contribution.
	}
\end{figure}

If  $A(P_\gamma),\, A(P_A)$ and $A(P_S)$  share the same  phases in short distance contributions, they  remain relative real in long distance ones, due to the same helicity. 
Hence, there are only 2 real parameters. 
The results of the  $\chi^2$ fitting are given by  
\begin{equation}
|p_A| = (5.5\pm 0.2 ) 10^{-2}\text{G}_F\text{GeV}^3\,,\,\,\,\,\,
p_S/p_A = 0.14 \pm 0.04\,,\,\,\,\,\chi^2/d.o.f. = 18/(15-2) \approx 1.4\,,
\end{equation}
where $d.o.f.$ stands for degree of freedom.
The 
reproduced
branching ratios with the available experimental data are given in Table~\MakeUppercase{\romannumeral 2}, while our predictions are shown in Table~\MakeUppercase{\romannumeral 3}, where we have used that
\begin{equation}\label{BRp}
{\cal B}^+_{exp} = {\cal B}_{exp} - {\cal B}^0_{exp} -{\cal B}^-_{exp}   =  (1-f_L){\cal B}_{exp}\,,
\end{equation}
for the data input~\cite{pdg} with the superscripts of $0$ and $\pm$ denoting the helicities,
and ${\cal B}^- = 0$.
In our results, the first uncertainties arise from the experimental input, and the second ones are the estimated  $10\%$ errors caused by the $SU(3)_F$ breaking.
The upper part of the table  contains the decays with $\overline{b}\to \overline{s}$, which are grouped according to the isospin group. The lower part 
  corresponds to those with  $\overline{b} \to \overline{d}$, which  are suppressed by $|V_{td}/V_{ts}|^2$.
We find that $p_A$ is about seven times larger than $p_S$, which is consistent with the OZI rule. 
Most of the reproduced branching ratios are consistent with the data.  However, there is a tension in  $B^0\to K^{*0} \omega$, as the theoretical value is almost twice larger than the experimental one, which could be a signal of NP.
\begin{table}[h]
	\captionsetup{justification=raggedright,
		singlelinecheck=false
	}
		\caption{Reproduced results in the SM, where the upper (lower) part  corresponds to $\overline{b} \to \overline{s}~(\overline{b}\to \overline{d})$, 
		and  the parametrized decay amplitudes of $O_{5,6}$ are given in the second column, in which the CKM matrix elements of $V_{ts,td}$  have not been shown explicitly for $\overline{b} \to (\overline{s}\,,\overline{d})$.} 
	\begin{tabular}{lccccc}
		\hline
channel &$SU(3)_F$& $10^6{\cal B}_{SU(3)}^+$&$10^6(1-f_L){\cal B}_{exp}$&
$\delta f_{exp}$\\
\hline
$ B^0  \to  K^{*0} \rho^{0} $& $ - p_{A}/\sqrt{2} $&$ 4.57 \pm 0.31 \pm 0.46$&$ 3.25\pm1.09$&$0.03\pm 0.08$\\
$ B^0  \to  K^{*+} \rho^{-} $&$ p_{A} $&$ 5.61 \pm 0.49 \pm 0.56$&$ 6.39\pm2.10$&\\
$ B^+  \to  K^{*+} \rho^{0} $&$ p_{A}/\sqrt{2} $&$ 1.59 \pm 0.19 \pm 0.16$&$ 1.01\pm0.6 0$&\\
$ B^+  \to  K^{*0} \rho^{+} $&$p_A$&$ 6.05 \pm 0.53 \pm 0.60$&$ 4.78\pm1.07$&\\
\hline
$ B^0  \to   K^{*0}\omega $&$  p_{A}/\sqrt{2} + \sqrt{2} p_{S} $&$1.14 \pm 0.17 \pm 0.11$&$ 0.62\pm0.27 $\\
$ B^+  \to  K^{*+}\omega $&$p_{A}/\sqrt{2}  + \sqrt{2} p_{S}$&$ 1.24 \pm 0.19 \pm 0.12 $&$ <4.4$&\\
\hline
$ B^0  \to  \phi K^{*0} $&$p_{A} + p_{S}$&$ 4.98 \pm 0.22 \pm 0.50 $&$ 5.03\pm0.30 $&$0.06 \pm 0.04$\\
$ B^+  \to  \phi K^{*+} $&$p_{A} + p_{S}$&$5.37 \pm 0.24 \pm 0.54 $&$ 5.00\pm1.12 $&$0.10\pm 0.11$\\
\hline
$ B_s  \to  \overline{K}^{*0} K^{*0} $&$ p_{A}$&$5.51 \pm 0.48 \pm 0.55$&$ 8.44\pm2.1 0$&$0.00\pm 0.24$\\
$ B_s  \to  \phi \phi $&$ 2p_{A} + 2 p_{S} $ &$9.60 \pm 0.43 \pm 0.96$&$ 11.63\pm0.96 $&$0.038 \pm 0.024$\\
\hline
\hline
$ B^0  \to  \rho^{0} \rho^{0} $&$p_A$&$  0.22 \pm 0.01 \pm 0.02 $&$ 0.28\pm0.10 $\\
$ B^0  \to  \rho^{-} \rho^{+} $&$p_A $&$0.25 \pm 0.02 \pm 0.02$&$ 0.28\pm0.55 $\\
$ B^0  \to  \overline{K}^{*0} K^{*0} $&$p_A$&$ 0.25 \pm 0.02 \pm 0.02$&$ 0.22\pm0.07 $\\
$ B^+  \to  \overline{K}^{*0} K^{*+} $&$p_A$&$  0.26 \pm 0.02 \pm 0.03 $&$ 0.16^{+0.20}_{-0.15} $\\
$ B^+  \to  \omega \rho^{+} $&$ \sqrt{2}p_A + \sqrt{2}p_S$&$0.36 \pm 0.02 \pm 0.04$&$ 1.59\pm0.98 $\\
$ B_s  \to  \phi \bar{K}^{*0} $&$ p_A + p_S$&$ 0.22 \pm 0.01 \pm 0.02$&$ 0.56\pm0.24 $&$0.07\pm 0.27$\\

		\hline
	\end{tabular}
	\label{fitting}
\end{table}

\begin{table}[h]
	\captionsetup{justification=raggedright,
		singlelinecheck=false
	}
		\caption{Predictions in the SM along with the parametrized decay amplitudes of $O_{5,6}$,
		 where the upper (lower) part  corresponds to $\overline{b} \to \overline{s}~(\overline{b}\to \overline{d})$\,, while the inequality holds in 95$\%$ confident level.
}
		\begin{tabular}[t]{lcccccc}
			\hline
			channel&$SU(3)_F$ & $10^6{\cal B}_{SU(3)}^+$\\
			\hline
			$ B_s  \to  \rho^{0} \rho^{0} $&0&$ 0 $\\
			$ B_s  \to  \rho^{-} \rho^{+} $&0&$ 0 $\\
			$ B_s  \to  K^{*-} K^{*+} $&$ p_{A} $&$ 5.51 \pm 0.48 \pm 0.55$\\
			$ B_s  \to  \omega \omega $&$0$&$ 0 $&$   $\\
			$ B_s  \to  \phi \omega $& $ \sqrt{2} p_{S} $ & $ 0.34 \pm 0.16 \pm 0.03 $\\
			$ B_s  \to  \phi \rho^0 $&0&$ 0.17 \pm 0.0 \pm 0.02 $\\
			\hline
			\hline
			$ B^+  \to  \phi \rho^{+} $&$p_S$&$<1.5 \times 10^{-3} $\\
			$ B^0  \to  K^{*-} K^{*+} $&$ 0 $&$ 0$\\
			$ B^0  \to  \omega \rho^{0} $&$  -p_A -p_S$&$0.26 \pm 0.01 \pm 0.02 $\\
			$ B^0  \to  \omega \omega $& $p_A + 2p_S$&$0.05 \pm 0.01 \pm 0.01$\\
			$ B^0  \to  \phi \rho^{0} $&$-p_S/\sqrt{2}$&$<0.7\times10^{-3}  $\\
			$ B^0  \to  \phi \omega $&$p_S/\sqrt{2}$&$<0.8\times10^{-3} $\\
			$ B^0  \to  \phi \phi $&0&$ 0 $\\
			$ B_s  \to  K^{*-} \rho^{+} $&$ p_A$&$ 0.24 \pm 0.02 \pm 0.02$\\
			$ B_s  \to  \overline{K}^{*0}\rho^{0} $&$-p_A/\sqrt{2}$&$ 0.20 \pm 0.01 \pm 0.02 $\\
			$ B_s  \to  \omega \bar{K}^{*0} $&$p_A/\sqrt{2} + \sqrt{2}p_S$&$ 0.05 \pm 0.01 \pm 0.01$\\
			\hline
		\end{tabular}
	\label{prediction}
\end{table}

It is interesting to note that
the effects of the charm-anticharm ($c\bar{c}$) rescattering, given as
\begin{eqnarray}
B\to \left(\begin{array}{ccc}
D_s^* D^* \\
D_sDK+X\\
 D_sDK+X\\
 \ldots
\end{array}\right)\to V_1V_2\,,
\end{eqnarray}
have not been considered yet, where ``$\ldots$'' represents other possible multiplicities.
However, excluding the cases that weak exchanging diagrams are permitted, in the $\overline{b} \to \overline{s} (\overline{d})$ transition, there would be at least one left-handed $\overline{s}(\overline{d})$ produced from the weak vertices, and therefore $A(H_-)$ is still suppressed. For instance, $B^0 \to K^{*0}\omega$ can not receive $A(H_-)$ from the $c\bar{c}$ rescattering in  the chiral limit.
Additionally, the relevant effective Hamiltonian for the rescattering, given as
\begin{equation}
{\cal H}_{eff}^c \propto V_{bc}^*V_{cs}(\overline{b}c)_L(\overline{c}s)_L +V_{bc}^*V_{cd} (\overline{b}c)_L(\overline{c}d)_L\,,
\end{equation}
has the same $SU(3)_F$ structure as the QCD penguin operators
 with the approximation of
 \begin{equation}
 V_{cd} /V_{cs} \approx V_{td}/V_{cd}\approx -\lambda \,,
 \end{equation}
 where
$\lambda = V_{us}$ is the parameter in the Wolfenstein parametrization for the CKM matrix~\cite{pdg}.
 In all, since the $c\bar{c}$ rescattering  approximately shares the same $SU(3)_F$ structure and $A(H_-)$ remains suppressed, the effects can be largely absorbed into  $A(P_A)$ and $A(P_S)$ by redefinition.

Let us examine CP violation. Since the transverse polarization  involves only one weak  phase in $V_{tb}^*V_{tf}$, the T violating effects are not expected to be observed. As a result,  we get
\begin{equation}\label{CPsingle1}
\Delta \phi _p = 0\,,
\end{equation}
which is well consistent with those in the  literature~\cite{Zou:2015iwa,FQCDBs,FQCDBu,SU3}.
Notice that there are two different equations concerning $\phi_p$. The first one is $\sin\phi_p\approx 0 $, argued 
from the helcitiy counting,
which therefore only constrains the strong phases. The second one is that $\Delta \phi_p= 0 $, originated from only one weak phase in the transverse amplitudes. Although their appearances seem similar, the mechanisms behind them are very different.
Consequently,  we arrive that 
\begin{equation}
\phi_\perp^B = \phi_\parallel^B\,,\quad \phi_\perp^{\overline{B}} =  \phi_\parallel^{\overline{B}}\pm \pi\,,
\end{equation}
which are well consistent with the previous theoretical works in Refs.~\cite{Zou:2015iwa,FQCDBs,FQCDBu,SU3,Beneke:2006hg,Wang:2017rmh}. 

In addition, we find that
$|A(H_+)|^2=|A(\overline{H}_{-})|^2$. Combining with $A(H_-) = A(\overline{H}_+) = 0$, we obtain 
\begin{equation}\label{CPsingle2}
A_{CP}^\perp =A_{CP}^\parallel =- A_{CP}^{dir}\,,
\end{equation}
where the CP asymmetries are defined by~\cite{pdg}
\begin{eqnarray}
A_{CP}^{dir} &=& \frac{\overline{\Gamma} - \Gamma}{\overline{\Gamma} + \Gamma}\,,\nonumber\\
A_{CP}^{\lambda}&
 =& \frac{\overline{f}_{\lambda} - f_{\lambda}}{\overline{f}_{\lambda} + f_{\lambda}}\,.
\end{eqnarray}
In Table \ref{CPdata},
we show the data related to these CP asymmetries~\cite{pdg,Krho,Aubert:2007ac}.
\begin{table}[t]
	\captionsetup{justification=raggedright,
		singlelinecheck=false
	}
	\caption{Data  for CP violating asymmetries, where $A_\perp^{CP}(B^+\to \phi K^{*+})$ 
		is obtained in Ref.~\cite{Aubert:2007ac}, while the other ones are taken from the PDG~\cite{pdg}.
	}
	\begin{tabular}{lcccc}
		\hline
		channel&$A_{\parallel}^{CP}$&$A_{\perp}^{CP}$&$A_{CP}^{dir}$\\ 
		\hline
		$B^0\to K^{*0}\rho^0 $& $0.19\pm 0.04$&$0.05\pm 0.04$&$-0.06\pm 0.09 $\\
		$B^0\to K^{*0}\omega $& $0.3\pm0.6$&$0.3\pm 0.9$&$0.45\pm 0.25$\\
		$B^0\to \phi K^{*0} $ &-&$-0.02\pm 0.06$&$0.00\pm 0.04$\\
		$B^+\to \phi K^{*+} $&-&$-0.22\pm 0.25$&$-0.01\pm 0.08$\\
		\hline
	\end{tabular}
\label{CPdata}
\end{table}

As seen in Table~\MakeUppercase{\romannumeral 1}, the central value of $\Delta \phi_p( B^0\to K^{*0}\omega)$ in the data shows  a huge discrepancy against Eq.~(\ref{CPsingle1}), which is sizable and not expected in the SM. 
One might look for explanation from the charm-anticharm rescattering, 
 which induces a  weak phase in $V_{cb}^*V_{cs}$. 
Nevertheless, it would be a futile attempt, since the relative weak phase between $V_{cb}^*V_{cs}$ and $V_{tb}^*V_{ts}$ vanishes in  the SM. 
 The another signal  could be seen in $B^0 \to K^{*0} \rho^0$.
Although the equality of $A_\perp^{CP}=-A^{dir}_{cp}$ holds well, there is a
  $2.5~\sigma$ deviation  between $A_\parallel^{CP}$ and $A_\perp^{CP}$,
which may arise from the right-handed current~\cite{Kagan:2004ia,Kagan:2004uw}.
However, the difference is small.
Interestingly, in $B^0\to K^{*0} \omega$, inconsistencies exist both in the decay branching ratio and T violating parameters.
If the refined experiments of $B^0 \to K^{*0}\omega$ in future confirm the data, it
would be a clear signal for NP.

Before ending the section, let us reexamine whether $O_{1,2}$ can be safely neglected in transverse amplitudes or not.  If they played a significant role in the transverse amplitudes,
  one would find an  enormous inconsistency in the $\overline{b}\to \overline{d}$ transition  in our approach, because of the hierarchy of $|V_{ud}/V_{td}|\gg |V_{us}/V_{ts}|$, {\it i.e.} the tree operators play  much more important roles in $\overline{b} \to \overline{d}$ compared to $\overline{b} \to \overline{s}$. 
Nevertheless, no such kind of phenomena has been found yet, and the systematical hierarchy
in Eq.~\eqref{SysH} holds as seen in Table~\ref{fitting}.
Particularly,  the data of $B^0\to \rho^0\rho^0, \overline{K}^{*0}K^{*0}$ fit  the theoretical expectations well.
As a result, the helicity flipping via $O_{1,2}^s$ is not capable of explaining the anomaly in ${\cal B}^+(B^0 \to K^{*0}\omega$). 
Finally, Eq.~(\ref{CPsingle1}) is uncontaminated by uncertainties from the experimental input,
and it is very unlikely for the SM to have $\Delta \phi_p (B^0 \to K^{*0} \omega)\sim0.8$. 
However, the uncertainties are still too large at this stage. Clearly, it is worth a refined experiment in future. 

 \section{Summary}
The eigenstates of T-odd scalar operators involving spins have been given for the first time.
From these eigenstates, we have explicitly constructed the T violating observables.
In addition, the relations between  the TVP asymmetries and the CP-odd phases have been  given consistently for the first time in Eq.~\eqref{consistent}.
  By counting the helicity flip, we have argued that $\sin(\phi_p)\approx 0 $, which matches the experimental data well.
Based on the helicity suppressions of $O_{1,2}$, we have unwrapped the interference between tree and penguin operators. Particularly,  $O_{1,2}$ could only give rise to the longitudinal polarization.
Our results for the transverse branching ratios have been represented in Tables~\ref{fitting} and \ref{prediction}. We have found that there is a tension between our approach and the experimental measurement in $B^0\to K^{*0}\omega$, in which the central value of our result is almost a factor of two larger than the data.

 Since $A(H_+)$ is dominated by  a single CP weak phase, a large CP violating effect in the transverse amplitudes is not expected.
 By deducing $\Delta \phi_p =0$,
we have shown that the central value of the experimental data,
  $\Delta \phi_p(B^0 \to K^{*0} \omega)=-0.84\pm 0.54 $~\cite{Krho}, can not be  explained within the SM.
  Interestingly, we have demonstrated inconsistencies in $B^0\to K^{*0}\omega$ for both transverse branching ratio and CP asymmetry. 
We recommend some dedicated  experiments in $B^0\to K^{*0}\omega$ to 
precisely measure these observables in order to claim that NP does exist.

 \appendix
 \section{Eigenstates of T-odd operators}
 
 In this Appendix, we calculate the eigenstates of $\hat{T}_i$ defined in Eq.~\eqref{definitionforTi} with the following commutation relation, given by~\cite{spinoperators}
\begin{equation}\label{spinoperator}
s_i L|\vec{p}=0, J_z = M\rangle = L J_i |\vec{p}=0, J_z = M\rangle \,,
\end{equation}
where $L$ is an arbitrary Lorentz boost.
For the explicit calculation, we take $\hat{T}_2$ as an example.
 Since $\hat{T}_2$  commutes with $\vec{J} \cdot \hat{p}_1$, the meson wave functions can be eigenstates of $\hat{T_2}$ and $\vec{J}\cdot\hat{p}_1$ simultaneously.
 We recall the canonical states for the two-particle system, given by 
 \begin{equation}
 |p\hat{z}, m_{s_1}, m_{s_2}\rangle = |p\hat{z}, m_{s_1}\rangle_1 \otimes |-p\hat{z}, m_{s_2}\rangle_2\,,
 \end{equation}
 where $m_{s_{1,2}}$ correspond to the eigenvalues of $(J_{1,2})_z$. We define the operator
 \begin{equation}\label{definitionofC2}
 \hat{C}_2 = s_{1x}s_{2y}- s_{1y}s_{2x} = \frac{i}{2}\left(
 s_1^+s_2^- - s_1^-s_2^+
 \right)
 \end{equation}
 with $s_i^\pm = s_{ix} \pm i s_{iy}$, which coincides with $\hat{T}_2$ on $|p\hat{z}\rangle$ and 
 offers a concrete form for the computation of the eigenstates.
 In the system with  vector and spin-$n$ mesons in the final states,  with Eq.~\eqref{spinoperator}, we find 
 \begin{eqnarray}\label{C2}
 &&|p\hat{z},J_z=0, \lambda_{t_2}=0\rangle = \frac{1}{\sqrt{2}}\left(|p\hat{z},m_{s_1}= 1,m_{s_2}=-1\rangle + |p\hat{z},m_{s_1}=-1,m_{s_2}=1\rangle\right)\,,\nonumber\\
 &&|p\hat{z},J_z=0, \lambda_{t_2}=\pm \sqrt{n^2+n}\rangle =
 \nonumber\\&& \frac{1}{2}\left( 
 |p\hat{z},m_{s_1}=1,m_{s_2}=-1\rangle -|p\hat{z},m_{s_1}=-1,m_{s_2}=1\rangle\mp \sqrt{2 }i |p\hat{z},m_{s_1}=0,m_{s_2}=0\rangle
 \right)\,,
 \end{eqnarray}
 where $\lambda_{t_2}$ is the eigenvalue of $\hat{T}_2$ and $\hat{C}_2$. Here, we have used $\vec{J}\cdot\hat{p}_1$  = $J_z$  in the subspace of $\hat{p}_1 = \hat{z}$.
 
 The states in Eq.~\eqref{C2} are eigenstates of $\hat{T}_2$, $\hat{p}_1$ and $\vec{J}\cdot \hat{p}_1$.
 To relate the states with the decays, we must construct them in a definite angular momentum to meet the initial condition, namely $J=J_z=0$, imposed by the $B$ meson. To this end, we utilize the projection operators in $SO(3)$ rotation group, given as \cite{GroupTheory,Jacob:1959at}
 \begin{equation}\label{projection}
 P^N_{JM}=\frac{2J+1}{8\pi^2}\int d\phi d\cos \theta d\psi U(\phi, \theta, \psi) D_J^\dagger(\phi,\theta,\psi)^N\,_M\,,
 \end{equation}
 where 
 $D_J$ is the Wigner D-matrix and 
 $U(\phi,\theta,\psi)=R_z(\phi)R_y(\theta)R_z(\psi)$ with $R_i$ the rotation operators in $\hat{i}$ direction.
 Here, $P^N_{JM}$ projects $|J, J_z=N \rangle$ on $|J, J_z=M\rangle$.
 After the projection, it remains the eigenstate of $\hat{T}_2$ since the scalar operators commute with $R$.
 
 By operating $P^0_{00}$ on Eq.~\eqref{C2}, we arrive the expected states. To compare the results with those in the literature, we write down them in terms of the  helicity states
 \begin{eqnarray}\label{lambdat2}
 |\lambda_{t_2}=0\rangle &=& |H_\parallel\rangle\,,\nonumber\\
 | \lambda_{t_2}=\pm \sqrt{n^2+n} \rangle &=& \frac{1}{\sqrt{2}}\left(
 |H_\perp \rangle \mp i |H_0\rangle
 \right)\,,
 \end{eqnarray}
with 
\begin{equation}\label{Hpm}
|H_{\perp,\parallel}\rangle = \frac{1}{\sqrt{2}}\left(
|H_+\rangle \mp |H_-\rangle
\right)\,,
\end{equation}
 where we have that $J=J_z=0$ in the both sides of the equations, and the subscripts in $H_{0,\pm}$ denote the helicities of the final state mesons.

 Notice that $\hat{T}_2$ and $\hat{T}_4$ are parity odd, whereas $\hat{T}_3$ parity even. All $T_i~(i=2,3,4)$ are designed as Hermitian operators.
 In 
 analogy to $\hat{C}_2$ in Eq.~\eqref{definitionofC2}, we write
 \begin{eqnarray}\label{C4}
 \hat{C}_3&=& s_{1z} \hat{C}_2 + \hat{C}_2 s_{1z}\,,\nonumber\\
 \hat{C}_4 &=&  \frac{i}{2}\left(
 s_1^+s_1^+s_2^-s_2^--s_1^-s_1^-s_2^+s_2^+
 \right)\,,
 \end{eqnarray}
 which coincide with $\hat{T}_3$ and $\hat{T}_4$, respectively, in the subspace of $\hat{p}_1= \hat{z}$. Consequently,
 the eigenvectors are given by
 \begin{eqnarray}\label{lambat3}
 &&|\lambda_{t_3} =0\rangle =| H_\perp  \rangle\,,\nonumber\\
 &&|\lambda_{t_3} =\pm \sqrt{n^2+n}\rangle = \frac{1}{\sqrt{2}}\left(
 |H_\parallel\rangle \mp i | H_0\rangle
 \right)\,,
 \end{eqnarray}
 and 
 \begin{eqnarray}\label{lambdat4}
 &&|\lambda_{t_4}=0\rangle = |H_0\rangle\,, \nonumber\\
 &&|\lambda_{t_4}=\pm (n^2 + n )\rangle = \frac{1}{\sqrt{2}}\left(
 |H_\parallel\rangle \pm i |H_\perp\rangle 
 \right)\,,
 \end{eqnarray}
 respectively, with $J=J_z=0$.
 
 Since we only concern spin-0 meson decays, it is sufficient to work on the subspace  with $\vec{J}\cdot\hat{p}_1=0$. As a result, one can substitute $\vec{s}_1\cdot \hat{p}_1$ by $-\vec{s}_2 \cdot \hat{p}_1$ in the definition of $\hat{T}$ without alternating the final results.

\end{document}